# EXPLODING NITROMETHANE IN SILICO, IN REAL TIME

**Eudes Eterno Fileti,[1] Vitaly V. Chaban,[2] and Oleg V. Prezhdo[3]**

[1] Instituto de Ciência e Tecnologia, Universidade Federal de São Paulo, 12231-280, São José dos Campos, SP, Brazil

[2] MEMPHYS - Center for Biomembrane Physics, Odense M, 5230, Kingdom of Denmark

[3] Department of Chemistry, University of Southern California, Los Angeles, CA, 90089, United States

**Abstract**. Nitromethane (NM) is widely applied in chemical technology as a solvent for extraction, cleaning and chemical synthesis. NM was considered safe for a long time, until a railroad tanker car exploded in 1958. We investigate detonation kinetics and reaction mechanisms in a variety of systems consisting of NM, molecular oxygen and water vapor. State-of-the-art reactive molecular dynamics allows us to simulate reactions in time-domain, as they occur in real life. High polarity of the NM molecule is shown to play an important role, driving the first exothermic step of the reaction. Presence of oxygen is important for faster oxidation, whereas its optimal concentration is in agreement with the proposed reaction mechanism. Addition of water (50 mol%) inhibits detonation; however, water does not prevent detonation entirely. The reported results provide important insights for improving applications of NM and preserving safety of industrial processes.

**E-mail for correspondence**: vvchaban@gmail.com; chaban@sdu.dk; fileti@gmail.com

**TOC image**

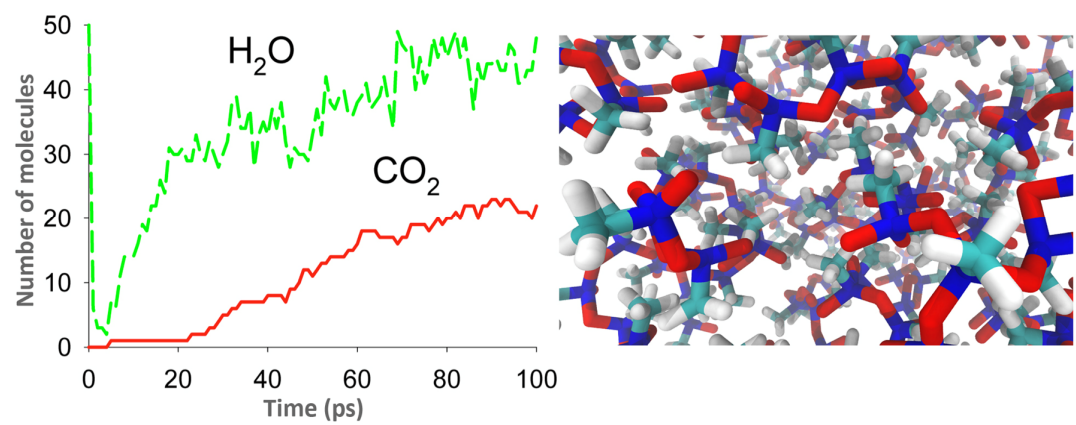

**Introduction**

Nitromethane (NM, $CH_3NO_2$) was developed in 1872 for a variety of chemical uses.[1] The versatility, functionality and efficiency of NM make it a great choice as a chemical additive for improving performance of various industrial processes.[2] An expedient source of nitrogen, NM is a highly functional intermediate in organic synthesis, used widely in manufacturing of pharmaceuticals,[3] as component of detergents, and as a chemical stabilizer in pesticides, explosives, fibers, and coatings.[1-2]

NM was not considered volatile in its pure form until 1950s, when numerous accidents occurred as it was being transported.[4] However, its use as fuel and explosive became quite common, mainly due to its low cost and safety. NM is utilized as a hobby fuel in model airplanes and racing cars, offering greater power and better performance. Explosive properties and combustion behavior of fuels based on NM have been studied at different levels, both theoretically and experimentally.[5-21] The processes involved in the shock initiation of detonation in pure NM have been studied by plate impact experiments under a steady one-dimensional strain for shock pressures ranging from 8.5 to 12 GPa.[19] The experimental results confirm the main steps of the classical homogeneous model,[22] but also show that the build-up process is more complex. When used as one of the additive components in propellants and explosives, liquid NM evaporates and mixes with air to form flammable gaseous nitromethane/air mixtures.[10, 18] The influence of the relative concentration of NM on the rate of increase of combustion pressure is relatively complicated.[10] Dilution of a liquid explosive, such nitrobenzene, with an inert, non-explosive liquid, such as methanol, benzene or water, can decrease the detonation velocity and, ultimately, destroy the ability of this mixture to detonate at ambient conditions. Expected from the energetic point of view, this situation has been investigated with nitrobenzene mixtures in recent years.[14] It has been concluded that the elusive behavior of the critical mixture cannot be completely described by simplified models of one-dimensional detonation theory.[14] Computational and numerical modeling have also been sporadically used.[11-13, 17] For instance, the decomposition mechanism of hot liquid NM has been investigated using density functional theory and reactive force field (ReaxFF) molecular dynamics (MD) simulations.[12, 23] Both studies elucidate molecular aspects of pure, liquid NM, revealing detailed chemical reactions that occur during detonation. Insights regarding the reaction mechanisms governing the initiation and detonation processes have been provided.

In this work, we investigate the detonation process for a variety of systems consisting of pure NM and NM mixed with molecular oxygen and water vapor. We employ state-of-the-art non-equilibrium reactive MD simulation, which allows us to observe reactions in real time. This methodology enables us to analyze the reactive mechanisms that drive such mixtures to explosion. We provide a general description of interactions of the mixture components, revealing the differences in the detonation processes at each composition.

**Methodology**

The MD simulations were performed using ReaxFF derived with quantum chemistry (QC).[24-27] This methodology was applied previously with success to address a number of complicated problems.[28-36] ReaxFF provides a nearly ab initio level of description of reactive potential surfaces for many-particle systems. The method treats all atoms in the system as separate interaction centers. The instantaneous point charge on each atom is determined by the electrostatic field due to all surrounding charges, supplemented by the second-order description of $dE/dq$, where $E$ is internal energy and $q$ is electrostatic charge on a given atom.[24-25] The interaction between two charges is written as a shielded Coulomb potential to guarantee correct behavior of covalently bonded atoms. The instantaneous valence force and interaction energy between two atoms are determined by the instantaneous bond order. The latter is determined by the instantaneous bond distance. These interaction energy functions are parametrized vs. QC energy scans involving all applicable types of bond-breaking processes. The bond order concept is used to define other valence interactions, such as bond, lone electron pair, valence angle, conjugation, and torsion angle energies. It is important for energy conservation and stability that all interaction terms smoothly decay to zero during bond dissociation. The pairwise van der Waals energy term describes short-range electron-electron repulsion, preserving atom size, and longer-range London attractive dispersion. Unlike non-reactive MD simulations, ReaxFF uses the van der Waals term for covalently bonded atoms, where it competes with a monotonically attractive bond term. Such an approach to chemical bonding requires a significant number of independent parameters, which can be obtained from QC energies. Bond dissociation, geometry distortion, electrostatic charges, infrared spectra, equations of state and condensed-phase structure are typically derived using an electronic structure method, such as density functional theory, to be consequently used in the ReaxFF parametrization. The works by van Duin, Goddard and coworkers[24-25] provide a more comprehensive description of the methodology used here.

The list of the simulated systems is provided in Table 1. Each system was simulated during 500 ps with an integration time-step of 0.1 fs. The explosion simulations were carried out in the constant energy ensemble (NVE), whereas the induced pressure was determined in the constant volume constant temperature (NVT) ensemble. We have also simulated constant pressure constant temperature (NPT) ensemble at 300 K to obtain liquid density of NM. Initial molecular configurations were generated using the PackMol[37] procedures to obtain system energies close to local minima. NVE simulations were started at 1000 K, and system temperature was monitored until 4000 K or slightly above depending on the system. 10-50 oxygen molecules were added to each system

(Figure 1) to represent atmosphere. Oxidants are expected to play an important role in thermodynamics and kinetics of the explosion reaction.

Table 1. Simulated systems: molecular compositions and representative parameters.

| # | # $CH_3NO_2$ | # $O_2$ | # atoms | Density, kg m$^{-3}$ | Time-step, fs | Duration, ps | T, K | Ensemble |
|---|---|---|---|---|---|---|---|---|
| 1 | 25 | 0 | 175 | 317 | 0.10 | 100 | 1000+ | NVE |
| 2 | 40 | 0 | 280 | 507 | 0.10 | 100 | 1000+ | NVE |
| 3 | 60 | 0 | 420 | 706 | 0.10 | 100 | 1000+ | NVE |
| 4 | 70 | 0 | 490 | 887 | 0.10 | 100 | 1000+ | NVE |
| 5 | 25 | 10 | 195 | 380 | 0.10 | 100 | 1000+ | NVE |
| 6 | 25 | 20 | 215 | 450 | 0.10 | 100 | 1000+ | NVE |
| 7 | 25 | 25 | 225 | 480 | 0.10 | 100 | 1000+ | NVE |
| 8 | 25 | 50 | 275 | 650 | 0.10 | 100 | 1000+ | NVE |
| 9 | 70 | 0 | 490 | 887 | 0.05 | 150 | 5000 | NVT |
| 10 | 25 | 25 | 195 | 380 | 0.05 | 150 | 5000 | NVT |
| 11 | 200 | 0 | 1400 | 1210 | 0.25 | 675 | 300 | NPT |
| 12 | 50 | 50$^{H2O}$ | 500 | 882 | 0.1 | 100 | 1000+ | NVE |

The algorithm of fragment recognition uses connection table and bond orders calculated at every time-step. The bond order cut-off used to identify molecular species is set to 0.3 for all bond types. Two fragments are considered separate molecules if all bond components, defined between them, exhibit orders smaller than 0.3. Note, that definition of a chemical bond is not unique, in principle. The selected value of the bond order cut-off influences the composition and concentration of intermediate products, but it does not influence the final (stable) products. Consequently, ReaxFF sporadically suggests existence of certain exotic molecules and fragments, which are not detected by any experimental technique, because of their transient nature and low stability. It is important to distinguish between bonded and non-bonded atom pairs to obtain translational kinetic energy, which is converted into temperature. The selected value (0.3) was tested in previous works, showing reliable and chemically sound results.

The non-equilibrium explosion dynamics was simulated in the following way. The potential energy of each system was minimized using the conjugate gradient algorithm for geometry optimization. The resulting configurations correspond to local energy minima in the absence of thermal motion. Afterwards, velocities corresponding to 1000 K, with respect to the initial number of covalent bonds, were assigned. The classical equations of motion were propagated using a very small integration time-step (0.0001 ps) conserving the total energy.

**Results and Discussions**

Unlike methane, NM is liquid at ambient temperature and pressure with density 1137 kg m$^{-3}$,[38] and enthalpy of vaporization 38.25 kJ mol$^{-1}$.[39] Van Duin and co-workers have investigated this liquid at ambient conditions, and at high pressures and temperatures. The reactive model provides density of 1111 kg m$^{-3}$. The difference of only 2% from the experimental value[12] suggests a high level of accuracy in the simulated physical chemical properties.

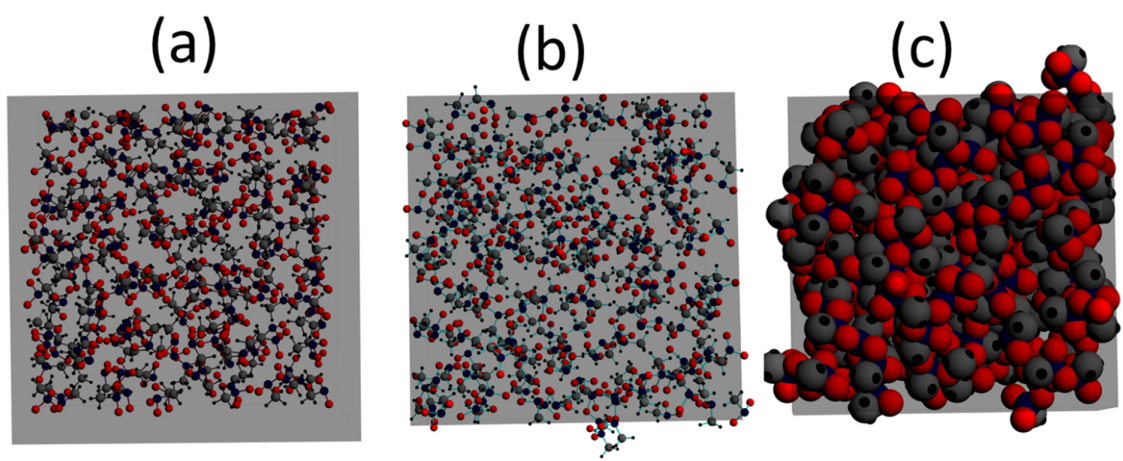

**Figure 1**. Molecular snapshots of the simulated liquid nitromethane at 300 K: (a) starting configuration; (b) final configuration with the density of 1210 kg m$^{-3}$; (c) final configuration (hard spheres representation). Oxygen atoms are red, nitrogen atoms are blue, carbon atoms are gray, hydrogen atoms are black.

Figure 2 shows temperature evolution in the pure NM systems at different densities. Initial heating up to 1000 K triggers the first reaction step (unimolecular decomposition). Afterwards, temperature grows drastically during the first 20 ps, irrespective of system density. Equilibrium in the lowest-density system (d=317 kg m$^{-3}$) is achieved during 40 ps. Higher density of the explosive brings higher temperature due to a larger number of chemical bonds per unit volume. Temperature exceeds 4000 K after the first 100 ps and continues to increase slowly. Interestingly, the difference between systems 2, 3, and 4 is smaller than their difference from system 1. If liberated heat is not dissipated efficiently, gaseous pressure in the reaction vessel increases by 500-2000 MPa (Figure 3).

Our result correlates reasonably with the recent data, where NM was simulated at high compression at 4000 K. According to van Duin and coworkers, application of 5000 MPa to NM results in a highly dense gas (d=1111 kg m$^{-3}$).[12] The density (Figure S1) and structure of liquid NM at 300 K are in satisfactory agreement with the experimental data and previous simulation results. NM is 20% denser than water at the same conditions, makes it a more energetic material. The density increase comes mainly from larger dipole moment of the NM molecule and, therefore, stronger electrostatic attraction energy.

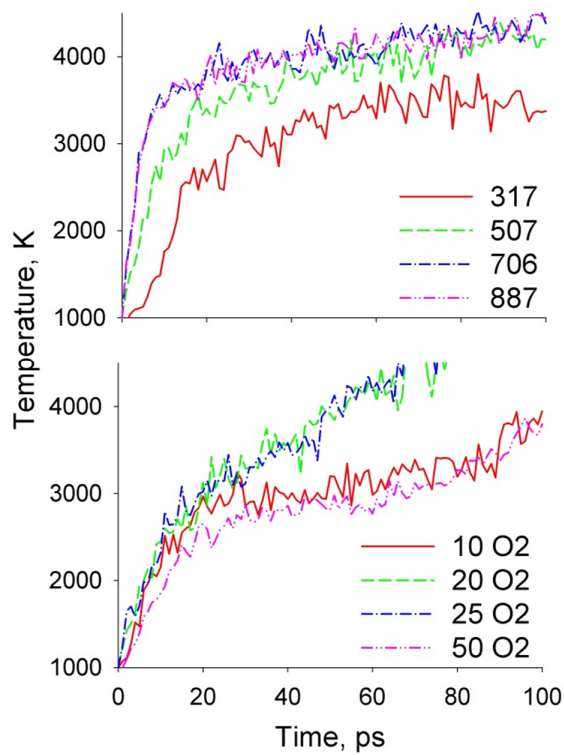

**Figure 2**. Temperature evolution in systems 1-8 computed in the constant total energy ensemble. (Top) systems of various density (see legends, in kg m$^{-3}$) containing only nitromethane. (Bottom) systems containing 25 nitromethane molecules and various quantities of oxygen molecules. The oxygen molecules are added to represent an atmosphere, which plays a significant role in the explosion process via oxidation.

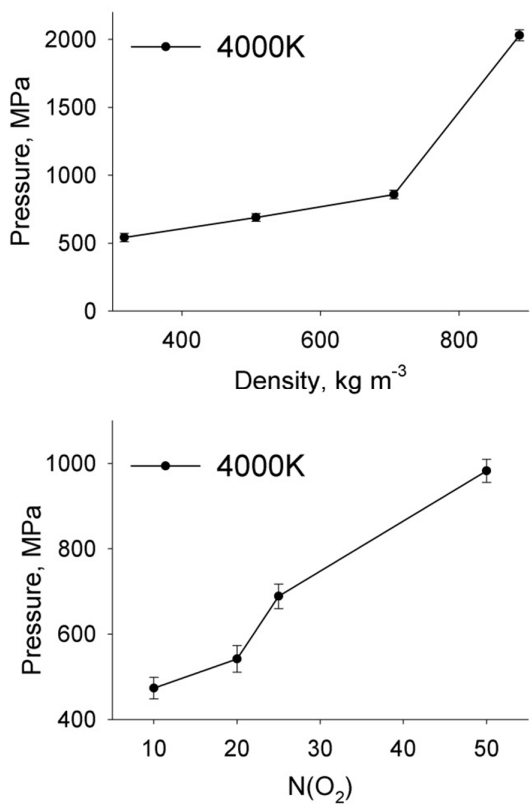

**Figure 3**. Pressure generated due to explosion in the simulated systems: (top) systems of various density (see legends, in kg m$^{-3}$) containing only nitromethane; (bottom) systems containing 25 nitromethane molecules and various quantities of oxygen molecules. Compare with Figure 2.

Atmospheric oxidants are expected to participate vigorously in the NM detonation, since metastable nitrogen and carbon containing particles need to transform into stable forms via oxidation. In particular, we envision two major reaction pathways:

$4CH_3NO_2 + 7O_2 = 4CO_2 + 4NO_2 + 6H_2O$     **4:7 ratio**     (I)
$4CH_3NO_2 + O_2 = 2CO_2 + 2N_2 + 6H_2O$     **4:1 ratio**     (II)

Formation of molecular nitrogen appears more probable, because N$_2$ is thermodynamically more stable at high temperatures. Formation of nitrogen dioxide is another possibility. These two competing reactions generate different amounts of CO$_2$. More importantly, they provide different numbers of newly synthesized gas molecules, which affect total pressure. That is, the particular pathway must depend on applied pressure and, therefore, on system density (Table 1). Another important observation is that different amounts of oxygen molecules are involved in the formation of $NO_2$ and $N_2$. Indeed, oxygen-poor systems follow mechanism II, as supported clearly by the lower temperature and pressure in these systems (Figures 2-3). Interestingly, the oxygen-richest system, $n(NM):n(O_2)=1:1$, also exhibits a slower reaction rate due to formation of a variety of oxygen containing metastable particles. Temperature of the $NM/O_2$ mixture grows rapidly during the first tens of picoseconds and continues a slower increase beyond 20 ps. Oxygen enrichment leads to density increase (Table 1) and, consequently, an increase in explosion pressure. The generated pressure is roughly proportional to oxygen content (Figure 3).

Figures 4-5 characterize kinetics of formation of the main reaction products, water and carbon dioxide. Figure S2 depicts a real-time decrease of the NM content. The NM decomposition rate clearly depends on system density, but it is not sensitive to $O_2$ concentration. Oxygen does not participate in the first stage of the explosion reaction. Molecular oxygen is a relatively weak oxidant. Oxidation at 1000 K is a slow process as compared to the timescale of the investigated detonation reactions. Interestingly, in the presence of oxygen, newly synthesized $CO_2$ and $H_2O$ molecules are more stable than in the case of the oxygen-free detonation. The quantity of oxygen atoms obtained from decomposition of NM is enough to give rise to $CO_2$ only, but hydrogen atoms cannot be oxidized (to generate water vapor). A possible pathway in this case is to form some compounds with nitrogen at high temperature and pressure, e.g. ammonia. However, the content of ammonia in the reaction by-products is negligible.

Figure 6 compares reaction products in the oxygen-free systems of various density and oxygen containing systems. More water molecules than $CO_2$ molecules are created in the $O_2$ deficient systems, while the trend inverts in the oxygen abundant systems. Modest amounts of $N_2$ are present in both kinds of systems. These three products are represented most, whereas a variety of other particles is formed in trace quantities ($NO_2$, $NH_3$, NHC, HNO, etc). If oxygen is lacking, the amount of such by-products exceeds 50%. An abundance of trace admixtures is also explained by a very high temperature (ca. 4000 K). The kinetic energy per degree of freedom exceeds the bond energies of many potentially stable products of NM decomposition, such as nitrogen oxides. In addition, the presence of both $CO_2$ and $H_2O$ in both kinds of systems suggests that formation of water vapor is much more energetically favorable than formation of molecular hydrogen or even the more reduced form of hydrogen, $NH_3$. Note, that initial oxidation state of hydrogen in this reaction is -1 (carbon hydride).

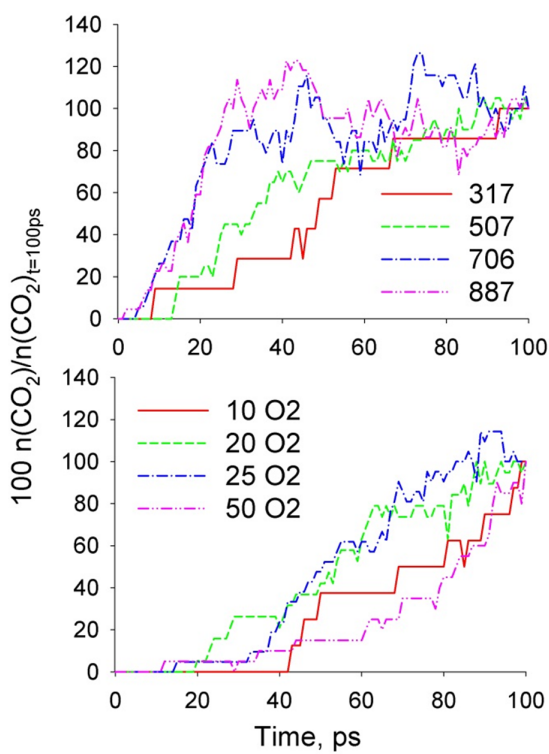

**Figure 4**. Kinetics of $CO_2$ formation during explosion. The $CO_2$ content is depicted as percentage of the final $CO_2$ content. At some points during the reaction, the content of $CO_2$ exceeds its final content, i.e. exceeds 100%. In the $O_2$ containing systems, $CO_2$ starts forming later than in the systems without oxygen.

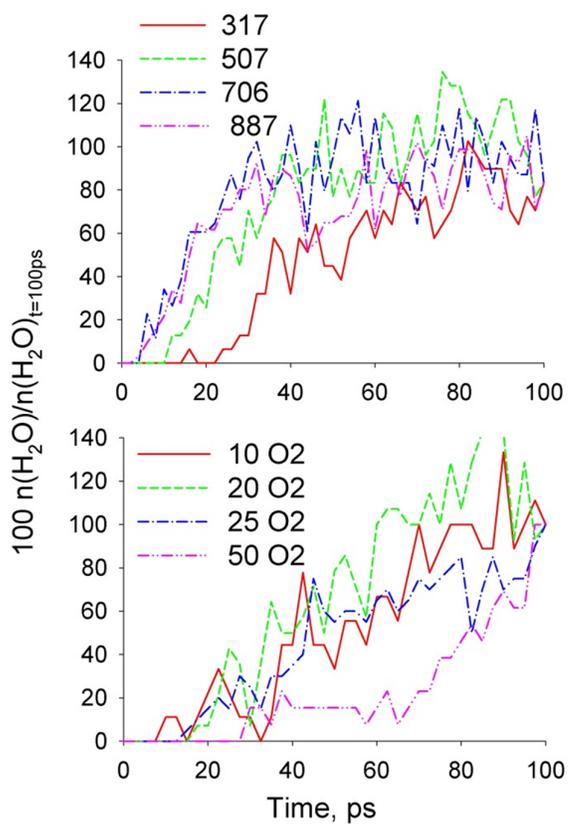

**Figure 5**. Formation of water vapor during explosion. The content of $CO_2$ molecules is depicted as percent of the final $CO_2$ content. At some points during the reaction, the content of $CO_2$ exceeds its final content, i.e. exceeds 100%. Water molecules are highly unstable at the simulated temperatures and their quantity fluctuates greatly.

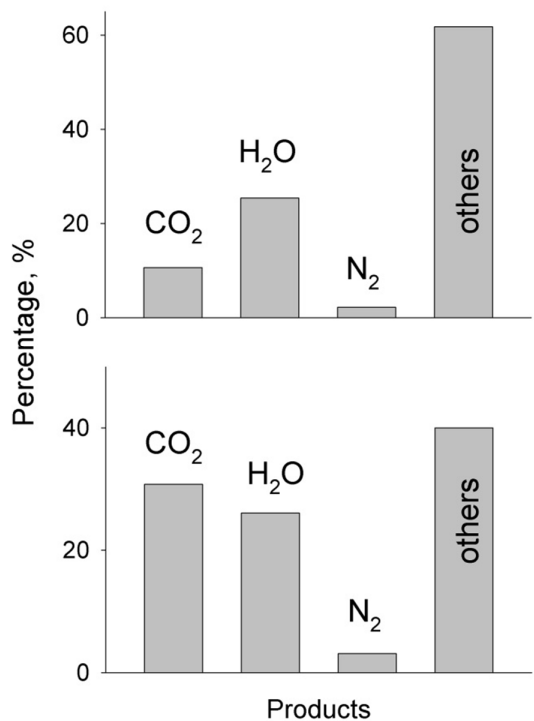

**Figure 6**. Final composition of the explosion products computed in the simulated systems 9-10, Table 1: (Top) in the system containing only $CH_3NO_2$ molecules (887 kg m$^{-3}$ density); (Bottom) in the system containing equimolar mixture of $CH_3NO_2$ and $O_2$.

NM is a popular solvent involved in a variety of industrial applications. Its safety, therefore, presents an important problem. To get insights into how explosivity of NM containing systems can be controlled, we simulated an equimolar NM-water mixture, system 12. The heat-initiated explosion was performed using the same computational procedure as for systems 1-8 (Figure 7). Water and NM exhibit a perfect mutual miscibility. A large amount of water does not block the explosion, but decreases somewhat the resulting temperature. Water induces strong hydrogen bonding with NM during the first stage that was interpreted by our algorithm as formation of a separate metastable molecule. Decrease in the water content during the first picoseconds illustrates this effect, Figure 6. The resulting compound, $CH_3–N(O)(OH)_2$, decomposes promptly providing water and carbon dioxide.

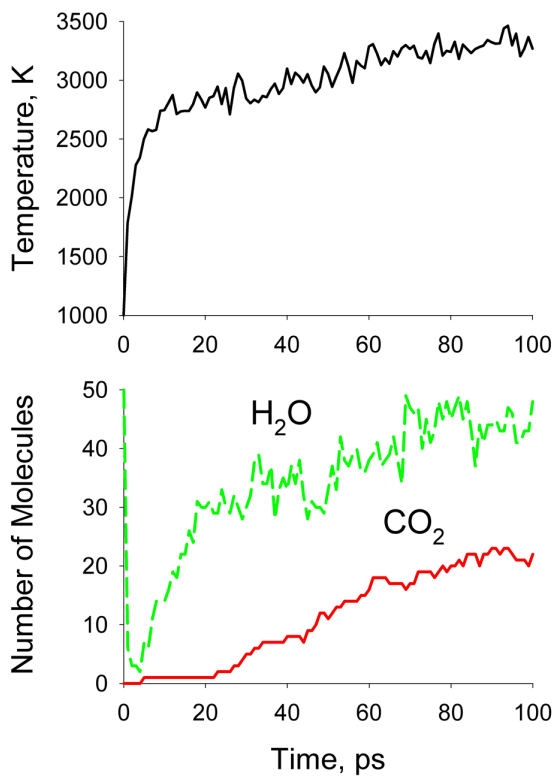

**Figure 7**. (Top) Temperature increase upon explosion of water-nitromethane equimolar mixture. (Bottom) Formation of water and carbon dioxide products during the first 100 ps of the reaction.

**Conclusions**

We have simulated a heat-initiated real-time explosion of NM in the absence and presence of molecular oxygen and water. While oxygen favors the detonation reaction, an excess of oxygen slows the reaction down. Such behavior is in agreement with the proposed reaction pathway. The gaseous products of the exothermic reaction generate significant pressures depending on the initial density of the reactants. Water impact on the detonation kinetics is explored for the first time, in particular, in regards to safety of NM as a popular solvent. We show that even 50 mol% addition of water does not block detonation in the case of heat-driven initiation. However, such a dilution of NM can be an important safety measure at room temperatures and ambient pressures.

## Acknowledgments

E. E. F. thanks Brazilian agencies FAPESP and CNPq for support. MEMPHYS is the Danish National Center of Excellence for Biomembrane Physics. The Center is supported by the Danish National Research Foundation. O.V.P. acknowledges grant CHE-1300118 from the US National Science Foundation. V.V.C. cordially thanks Prof. Adri van Duin for important discussions regarding reactive molecular dynamics methodology and force field development. Dr. Fedor Goumans (Scientific Computing & Modeling, Amsterdam) has provided invaluable assistance in visualization of the simulation results.

## Supporting Information Available

Figures S1 and S2 referenced in the manuscript can be found in Supporting Information. This information is available free of charge via the Internet at http://pubs.acs.org.